\documentstyle[preprint,aps]{revtex}
\newcommand{\beq}{\begin{equation}}
\newcommand{\eeq}{\end{equation}}
\newcommand{\beqa}{\begin{eqnarray}}
\newcommand{\eeqa}{\end{eqnarray}}
\newcommand{\ba}{\begin{array}}
\newcommand{\ea}{\end{array}}

\begin{document}
\draft


\widetext 

\title{Pathological Behavior in the Spectral Statistics \\
of the Asymmetric Rotor Model} 
\author{V.R. Manfredi$^{1}$ and L. Salasnich$^{2}$} 
\address{
$^{1}$Dipartimento di Fisica "G. Galilei", Universit\`a di Padova, \\
Istituto Nazionale di Fisica Nucleare, Sezione di Padova, \\
Via Marzolo 8, 35131 Padova, Italy, \\
$^{2}$Istituto Nazionale per la Fisica della Materia, 
Unit\`a di Milano Universit\`a,\\ 
Dipartimento di Fisica, Universit\`a di Milano, \\ 
Via Celoria 16, 20133 Milano, Italy} 

\maketitle

\begin{abstract} 
The aim of this work is to study the spectral statistics 
of the asymmetric rotor model 
(triaxial rigid rotator). 
The asymmetric top is classically integrable and, according 
to the Berry-Tabor theory, its spectral statistics should be 
Poissonian. Surprisingly, our numerical 
results show that the nearest neighbor spacing distribution 
$P(s)$ and the spectral rigidity $\Delta_3(L)$ 
do not follow Poisson statistics. 
In particular, $P(s)$ shows a sharp peak at 
$s=1$ while $\Delta_3(L)$ for small values of $L$ 
follows the Poissonian predictions and asymptotically 
it shows large fluctuations around its mean value. 
Finally, we analyze the information entropy, 
which shows a dissolution of quantum 
numbers by breaking the axial symmetry of the rigid rotator. 

\end{abstract}

\vskip 0.5cm 



\narrowtext

\section{introduction}

In the semiclassical limit [1,2] there is 
a clear connection between the behavior of classical systems 
(regular or chaotic) and the corresponding quantal ones. 
For quantal systems, corresponding to classical 
regular systems, the spectral statistics ($P(s)$ and $\Delta_3(L)$) 
follow the Poisson ensemble, while for systems corresponding 
to chaotic ones the Wigner ensemble is followed 
(see, for example, [3] and references therein). 
\par
Nevertheless, some exceptions are known. 
The most famous case is perhaps the harmonic oscillator one, discussed in great detail in references [4,5]. 
It has been also found that low-energy spectral 
statistics of higher-dimensional separable 
Hamiltonian systems can show the level repulsion typical for 
chaotic systems. Especially critical in this sense are systems 
close to the harmonic oscillators and to rectangular wells [6]. 
\par 
The aim of this paper is to discuss another pathological case: 
the classically integrable triaxial rotator model (see, for 
instance, ref. [7]). Incidentally, this model has been used 
very often in the description of the low-lying states of the 
even-even atomic nuclei [8]. 
\par 
The asymmetric top described by the rotor model 
is a classically integrable system, but 
an analytical formula, as a function of quantum numbers, 
for its energy spectrum is not known. 
Nevertheless, numerical results can be obtained. 
By following the Landau approach [9], the Hamiltonian operator 
is split into 4 submatrices, corresponding to different symmetry classes. 
Each truncated submatrix is numerically diagonalized. 
Finally, the nearest-neighbor spacing distribution $P(s)$ 
and the spectral rigidity $\Delta_3(L)$ are calculated. 
Surprisingly, the spectral statistics of energy levels do not 
follow the predictions of Poisson statistics. 

\section{The Asymmetric Rotor Model}

Let us consider a system of coordinates with axes along 
the three principal axes of intertia of the top, and rotating 
with it. The classical Hamiltonian $H$ of the top is given by 
\beq 
{\hat H} = {1 \over 2} 
\left( a J_1^2 + b J_2^2  + c J_3^2 \right) \; , 
\eeq 
where ${\bf J}=(J_1,J_2,J_3)$ is the angular momentum of the 
rotation and $a=1/I_1$, $b=1/I_2$, $c=1/I_3$ are three 
parameters such that $I_1$, $I_2$ and $I_3$ are 
the principal momenta of intertia of the top. 
The Hamiltonian is classically integrable 
and its action variables are precisely 
the three components $J_s$, $s=1,2,3$, of the 
angular momentum [7]. 
\par 
The quantum Hamiltonian ${\hat H}$ is 
obtained by replacing the components of the 
angular momentum, in the classical expression 
of the energy, by the corresponding quantum operators 
${\hat J}_1$, ${\hat J}_2$ and ${\hat J}_3$.  
The commutation rules for the operators of the angular momentum 
components in the rotating system of coordinates are given by 
\beq 
{\hat J}_r {\hat J}_s - {\hat J}_s {\hat J}_r 
= - i\hbar \; \epsilon_{rst} \; {\hat J}_t \; , 
\eeq 
where $\epsilon_{rst}$ is the Ricci tensor and $r,s,t=1,2,3$. 
Note that these commutation rules  
differ from those in the fixed system in the sign on the 
right-hand side [9]. 
\par 
As usual, the two operators 
${\hat J}^2={\hat J}_1^2+{\hat J}_2^2+{\hat J}_3^2$ 
and ${\hat J}_3$ are simultaneously diagonalized on the basis of 
eigenstates $|J,k\rangle$ with integer eigenvalues $J$ and $k$ 
($k=-J,-J+1,...,J-1,J$), respectively. 
The non-zero matrix elements of the quantum Hamiltonian ${\hat H}$ 
in the basis $|J,k\rangle$ are given by 
\beq 
\langle J,k|{\hat H}|J,k \rangle = {\hbar^2 \over 4} (a+b) 
(J(J+1)-k^2) + {\hbar^2 \over 2} c k^2 \; , 
\eeq
$$
\langle J,k|{\hat H}|J,k+2 \rangle = \langle J,k+2|H|J,k \rangle = 
$$ 
\beq
= {\hbar^2 \over 8} (a-b)\sqrt{ (J-k)(J-k-1)(J+k+1)(J+k+2) } \; . 
\eeq 
The quantum Hamiltonian ${\hat H}$ has matrix elements only 
for transitions with $k\to k$ or $k\pm2$. The absence of matrix 
elements for transitions between states with even and odd $k$ 
has the result that the matrix of degree $2J+1$ is the 
direct product of two matrices of degrees 
$J$ and $J+1$. One of these contains matrix elements for transitions 
between states with even $k$, and the other contains those 
for transitions between states with odd $k$ [9]. 
\par 
It is useful to introduce a new basis, given by 
\par 
$$ 
|J,k;S\rangle 
= {1\over \sqrt{2}} \left( |J,k\rangle + |J,-k\rangle \right) \; , 
\;\;\;\;\; |J,0,S\rangle =  |J,0\rangle \; , 
$$ 
\beq 
|J,k;A\rangle 
= {1\over \sqrt{2}} \left( |J,k\rangle - |J,-k\rangle \right) 
\; , \;\;\;\;\; k\neq 0 \; . 
\eeq 
By using this new basis, the total Hamiltonian matrix 
is decomposed in the direct product of $4$ submatrices 
by considering the parity of the quantum 
number $k$: even (E) or odd (O), and the symmetry of the state: 
symmetric (S) or anti-symmetric (A). 
So the submatrices are labelled as follow: (E,S), (E,A), (O,S), (O,A). 
These are the classes of symmetry of the system. 
In Table 1 we show the dimension of each submatrix for a fixed $J$. 
\par
The matrix elements of the Hamiltonian ${\hat H}$ in the 
new basis, with respect to the old basis, are given by 
\beq
\langle J,k,S |{\hat H}|J,k,S \rangle = 
\langle J,k,A |{\hat H}|J,k,A \rangle 
= \langle J,k|{\hat H}|J,k \rangle 
\; , \;\;\;\;\; k\neq 1 
\eeq
\beq 
\langle J,1,S |{\hat H}|J,1,S \rangle = 
\langle J,1|{\hat H}|J,1 \rangle \; + \; 
\langle J,1|{\hat H}|J,-1\rangle 
\eeq
\beq 
\langle J,1,A |{\hat H}|J,1,A \rangle = 
\langle J,1|{\hat H}|J,1 \rangle \; - \; 
\langle J,1|{\hat H}|J,-1\rangle 
\eeq
\beq
\langle J,k,S |{\hat H}|J,k+2,S \rangle = 
\langle J,k,A |{\hat H}|J,k+2,A \rangle 
= \langle J,k|{\hat H}|J,k+2 \rangle 
\; , \;\;\;\;\; k\neq 0 
\eeq
\beq
\langle J,0,S |{\hat H}|J,2,S \rangle = 
\sqrt{2} \langle J,0|{\hat H}|J,2 \rangle 
\; , \;\;\;\;\; k\neq 0 
\eeq
We calculate the eigenvalues of each submatrix 
for different values of $J$ using a fast implementation, 
in double precision, 
of the Lanczos algorithm with a LAPAC code [10]. 
In Figure 1 we plot the density of levels $\rho(E)$ 
of each submatrix of ${\hat H}$ and $J=1000$. 
The results show that the density of levels is practically 
the same for the four classes. $\rho(E)$ displays a high peak 
at the left-center of the energy interval and a long tail 
for large energy values. 

\section{Spectral Statistics} 

As previously discussed, according to the Berry-Tabor theory 
[11,12], given a classical integrable Hamiltonian that, 
written in action variables $J_r$, 
satisfies the condition 
\beq  
\left| {\partial^2 H\over \partial J_r \partial J_s} 
\right| \neq 0 \; , 
\eeq 
then, in the semiclassical limit, 
its spectral statistics should follow the Poisson statistics. 
Note that a system of linear harmonic oscillators, whose 
Hamiltonian is given by $H={\bf \omega}\cdot {\bf I}$, 
does not satisfy the previous condition. 
In fact, a system of linear harmonic oscillators is integrable 
but it does not follow Poissonian statistics [4,5]. 
\par 
The triaxial rigid rotator is integrable and satisfies 
the Berry-Tabor condition (11). Thus, one expects that 
the spectral statistics of the quantized rigid rotator 
should be Poissonian. We shall show that is not the case. 
\par 
In general, various statistics may be used to show the local correlations 
of the energy levels but the most used spectral statistics 
are $P(s)$ and $\Delta_3(L)$. $P(s)$ is 
the distribution of nearest-neighbor spacings 
$s_i=({\tilde E}_{i+1}-{\tilde E}_i)$ 
of the unfolded levels ${\tilde E}_i$. 
It is obtained by accumulating the number of spacings that lie within 
the bin $(s,s+\Delta s)$ and then normalizing $P(s)$ to unit. 
As shown by Berry and Tabor [11,12], 
for quantum systems whose classical analogs are integrable, 
$P(s)$ is expected to follow the Poisson distribution 
\beq 
P(s)=\exp{(-s)} \; . 
\eeq 
The statistic $\Delta_3(L)$ is defined, for a fixed interval 
$(-L/2,L/2)$, as the least-square deviation of the staircase 
function $N(E)$ from the best straight line fitting it: 
$$
\Delta_{3}(L)={1\over L}\min_{A,B}\int_{-L/2}^{L/2}[N(E)-AE-B]^2 dE \; , 
$$
where $N(E)$ is the number of levels between E and zero for positive 
energy, between $-E$ and zero for negative energy. 
The $\Delta_{3}(L)$ 
statistic provides a measure of the degree of rigidity of the 
spectrum: for a given interval L, the smaller $\Delta_{3}(L)$ is, 
the stronger is the rigidity, signifying the long-range 
correlations between levels. For this statistic 
the Poissonian prediction is 
\beq 
\Delta_3(L)= {L\over 15} \; . 
\eeq 
It is useful to remember that Berry, on the basis 
of the Gutwiller semiclassical formula for the density of states, 
has shown that $\Delta_3(L)$ deviates from the universal 
Poissonian predictions for large $L$: 
$\Delta_3(L)$ should saturate to an asymptotic value 
performing damped oscillations [13]. 
\par 
In Figure 2 the spectral statistic 
$P(s)$ is plotted for the four 
submatrices of ${\hat H}$ and $J=1000$.  
Note that the level spectrum is mapped into unfolded levels 
with quasi-uniform level density 
by using a standard procedure described in [14]. 
As expected from the previous analysis of density of levels, 
$P(s)$ is practically the same for the four classes of symmetry. 
Moreover, $P(s)$ has a pathological behavior: a peak near 
$s=1$ and nothing elsewhere. 
Compared to $P(s)$, the spectral rigidity $\Delta_3(L)$ 
is less pathological. As shown in Figure 3, $\Delta_3(L)$ 
follows quite well the Poisson prediction $\Delta_3(L)=L/15$ 
for small $L$ but for larger values of $L$ 
it gets a constant mean value with 
fluctuations around this mean value. These fluctuations 
becomes very large by increasing $L$, in contrast 
with the Berry prediction [13]. 
\par 
The behavior of the density of levels $\rho(E)$ and 
of the spectral statistics $P(s)$ and $\Delta_3(L)$ 
does not change by changing the matrix dimension, 
namely the quantum number $J$. In Figure 4 we plot 
the density of levels and the spectral statistics 
for $J=2000$ and $J=4000$. 
\par
The results shown in the first four figures have been obtained 
with $a=1$, $b=\sqrt{2}$ and $\sqrt{5}$, in such a way 
that the rotor is triaxial. 
It is interesting to see what happens if one 
changes the deformation parameters $a$, $b$ and $c$, 
studying the transition 
from axial symmetry to triaxial symmetry. 
To do so, we take fixed $b$ and $c$ and modify $a$. 
In Figure 5 we plot the density of levels $\rho(E)$ 
for six values of $a$ ranging from $a=\sqrt{2}$ to 
$a=0$. The case $a=b=\sqrt{2}$ 
correspond to the axial symmetric one. 
The density of levels $\rho(E)$ is strongly modified by changing 
the parameter $a$, i.e. breaking the axial symmetry, 
but the spectral statistics are not, 
as shown by Figure 6 for the $P(s)$ distribution. 
\par 
To conclude this section, we discuss 
another statistical quantity that has been proposed 
to study quantum chaos: the information entropy $S(E)$ 
of the eigenvector $|E>$ associated to the eigenvalue 
$E$ of the the Hamiltonian operator ${\hat H}$ [15]. 
Given a generic basis set $\{ |i> \}$, 
the eigenvector $|E>$ can be written as: 
\beq 
|E> = \sum_i c_i |i> \; , 
\eeq 
where $c_i$ are the probability amplitudes. Then, 
the information entropy of the eigenvector $|E>$ 
with respect to the basis set $\{|i>\}$ is defined as 
\beq 
S(E) = -\sum_i |c_i|^2 \ln{|c_i|^2} \; . 
\eeq
The idea is that, just as in the classical 
theory a dissolution of integrability (with the KAM mechanism) simply 
means the onset of chaotic motion, in quantum systems a dissolution 
of quantum numbers may indicate the onset of 
quantum chaos (see also [16]). 
In figure 7 we show the information entropy $S(E)$ 
of the eigenvectors of the Hamiltonian matrix of symmetry class 
(E,S) with respect to the axial symmetric 
basis set $|J,k,S>$, calculated for different values of the 
deformation parameter $a$. As expected, 
if the system has axial symmetry ($a=\sqrt{2}$) 
then the information entropy $S(E)$ is everywhere zero. 
By deforming the system, i.e. breaking the axial symmetry, 
$S(E)$ becomes positive and it is larger in the central 
part of the energy spectrum. 
It is important to stress that in our system the dissolution 
of quantum numbers shown in Figure 7 does not have a classical 
analog because the classical Hamiltonian is always integrable. 
Thus, in our case, a large $S(E)$ simply means a fully 
broken axial symmetry of the rigid rotor. 

\section*{Conclusions} 

The main conclusion of this paper 
is that the asymmetric rotor is, like the harmonic 
oscillator, another pathological case with respect 
to the classical-quantum correspondence 
between integrability and Poisson statistics. 
In our opinion, the pathology of the asymmetric rotor model 
is more interesting because, unlike the harmonic oscillator, 
the asymmetric rotor satisfies the conditions 
of the Berry-Tabor theory. 
The presence of hidden symmetries could explain 
the pathological behavior of spectral statistics 
but such symmetries have not yet been identified. 
For the sake of completeness we remember that, as stressed 
by Rau [17], in classical mechanics, the asymmetric rotor and the 
nonlinear pendulum are intimately linked and form the basis 
for many studies of nonlinear dynamics. 
Finally, we have shown that the information entropy of 
eivenvectors with the respect to the axial symmetric basis set 
gives a clear signature of the breaking of axial symmetry of the 
rigid rotator, but the rigid rotator is always 
classically integrable. 

\vskip 0.5cm 

V.R.M. is greately indebted to M.V. Berry and S. Graffi 
for enlightening discussions. 
L.S. acknowledges M. Robnik and J.M.G. Gomez 
for fruitful conversations. 

\newpage

\section*{Tables} 

\vskip 0.5cm

\begin{center}
\begin{tabular}{|ccccc|} \hline 
$$ & $(E,S)$ & $(E,A)$ & $(O,S)$ & $(O,A)$ \\ 
\hline 
$J$ even & ${J\over 2}+1$ & ${J\over 2}$   & ${J\over 2}$   & ${J\over 2}$ \\ 
$J$ odd  & ${J-1\over 2}$ & ${J+1\over 2}$ & ${J+1\over 2}$ & ${J+1\over 2}$ \\ 
\hline 
\end{tabular} 
\end{center} 

\vskip 0.3 truecm 
{\bf Table 1}. Number of states in each submatrix of the 
asymmetrical top Hamiltonian for a fixed $J$. 

\vskip 0.5cm 

\newpage

\section*{Figure Captions} 

{\bf Figure 1}: Density of levels $\rho(E)$ of the 
four classes of symmetry with $J=1000$. 
Parameters: $a=1$, $b=\sqrt{2}$, 
$c=\sqrt{5}$ and $\hbar=1$. 

{\bf Figure 2}: Nearest neighbor spacing distribution 
$P(s)$ of the four classes of symmetry with $J=1000$. 
The dashed line is the Poisson prediction $P(s)=\exp{(-s)}$. 
Parameters: $a=1$, $b=\sqrt{2}$, 
$c=\sqrt{5}$ and $\hbar=1$. 

{\bf Figure 3}: Spectral rigidity $\Delta_3(L)$ of the 
four classes of symmetry with $J=1000$. 
The dashed line is the Poisson prediction $\Delta_3(L)=L/15$. 
Parameters: $a=1$, $b=\sqrt{2}$, $c=\sqrt{5}$ and $\hbar=1$. 

{\bf Figure 4}: Density of levels $\rho(E)$, 
Nearest neighbor spacing distribution 
$P(s)$ and spectral rigidity $\Delta_3(L)$ 
of the $(E,S)$ class of symmetry, with 
$J=2000$ (top) and $J=4000$ (bottom). 
Dashed lines are Poisson predictions. 
Parameters: $a=1$, $b=\sqrt{2}$, 
$c=\sqrt{5}$ and $\hbar=1$. 

{\bf Figure 5}: Density of levels $\rho(E)$  
of the $(E,S)$ class of symmetry, with $J=1000$. 
Parameters: $b=\sqrt{2}$, 
$c=\sqrt{5}$ and $\hbar=1$. 
Different values of the deformation parameter: 
(a) $a=\sqrt{2}$, (b) $a=\sqrt{1.9}$, 
(c) $a=\sqrt{1.5}$, (d) $a=1$, 
(e) $a=\sqrt{0.5}$, (f) $a=0$. 

{\bf Figure 6}: Nearest neighbor spacing distribution 
$P(s)$  of the $(E,S)$ class of symmetry, with $J=1000$. 
Parameters: $b=\sqrt{2}$, 
$c=\sqrt{5}$ and $\hbar=1$. 
Different values of the deformation parameter: 
(a) $a=\sqrt{2}$, (b) $a=\sqrt{1.9}$, 
(c) $a=\sqrt{1.5}$, (d) $a=1$, 
(e) $a=\sqrt{0.5}$, (f) $a=0$. 

{\bf Figure 7}: Information entropy $S(E)$  
of the $(E,S)$ class of symmetry, with $J=1000$. 
Parameters: $b=\sqrt{2}$, 
$c=\sqrt{5}$ and $\hbar=1$. 
Different values of the deformation parameter: 
(a) $a=\sqrt{2}$, (b) $a=\sqrt{1.9}$, 
(c) $a=\sqrt{1.5}$, (d) $a=1$, 
(e) $a=\sqrt{0.5}$, (f) $a=0$. 

\newpage 

\section*{References} 

[1] V.P. Maslov and M.V. Fedoriuk, {\it Semiclassical Approximation 
in Quantum Mechanics} (Reidel Publishing Company, 1981). 

[2] A.B. Migdal, {\it Qualitative Methods in Quantum Theory}  
(Benjamin, 1997). 

[3] V.R. Manfredi and L. Salasnich, 
Int. J. Mod. Phys. B {\bf 13}, 2343 (1999). 

[4] A. Pandey, O. Bohigas and M.J. Giannoni: J. Phys. A: Math. Gen. 
{\bf 22}, 4083 (1989). 

[5] A. Pandey and R. Ramaswamy: Phys. Rev. A {\bf 43}, 4237 (1991). 

[6] S. Drozdz and J. Speth: Phys. Rev. Lett. {\bf 67}, 529 (1991); 
{\bf 68}, 3109 (1992). 

[7] H. Goldstein, {\it Classical Mechanics} 
(Addision Wesley, Reading, 1980). 

[8] J.M. Eisenberg and W. Greiner, {\it Nuclear Models}, vol. 1 
(North Hollnd, Amsterdam, 1975). 

[9] L. Landau and E. Lifshitz, {\it Course in Theoretical Physics}, 
vol. 3: Quantum Mechanics (Pergamon, 1977).  

[10] LAPAC Fortran Library, Linear Algebra Package, NAG Ltd 2001. 

[11] M.V. Berry and M. Tabor, Proc. Roy. Soc. Lond. A {\bf 356}, 375 (1977); M.V. Berry, Annals of Phys. {\bf 131}, 163 (1981). 

[12] M. Tabor {\it Chaos and Integrability in Nonlinear Dynamics} 
(Wiley, New York, 1989). 

[13] M. Berry, Proc. Roy. Soc. Lond. A {\bf 400}, 229 (1985). 

[14] V.R. Manfredi, Lett. Nuovo Cimento {\bf 40}, 135 (1984). 

[15] J. Reichl, Europhys. Lett. {\bf 6}, 669 (1988). 

[16] F. Sakata {\it et al.}, Nucl. Phys. A {\bf 519}, 93c (1990). 

[17] A.R.P. Rau, Rev. Mod. Phys. {\bf 64}, 623 (1992). 

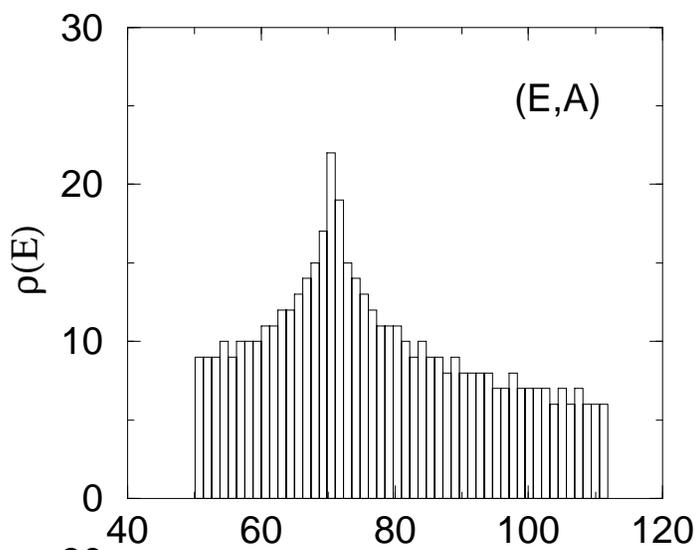
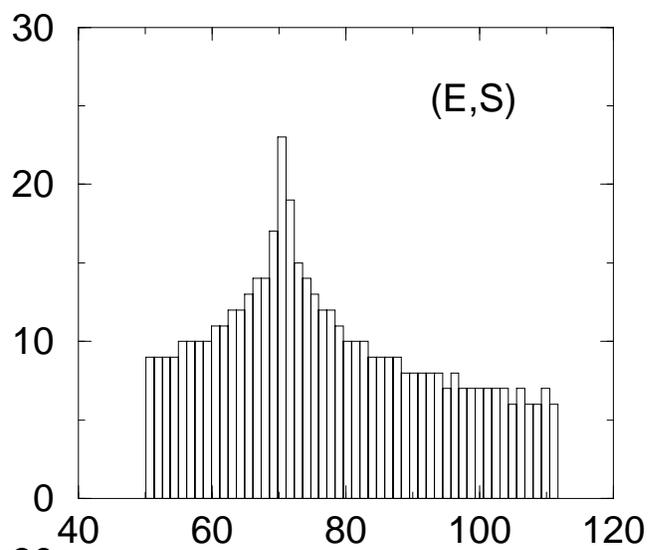
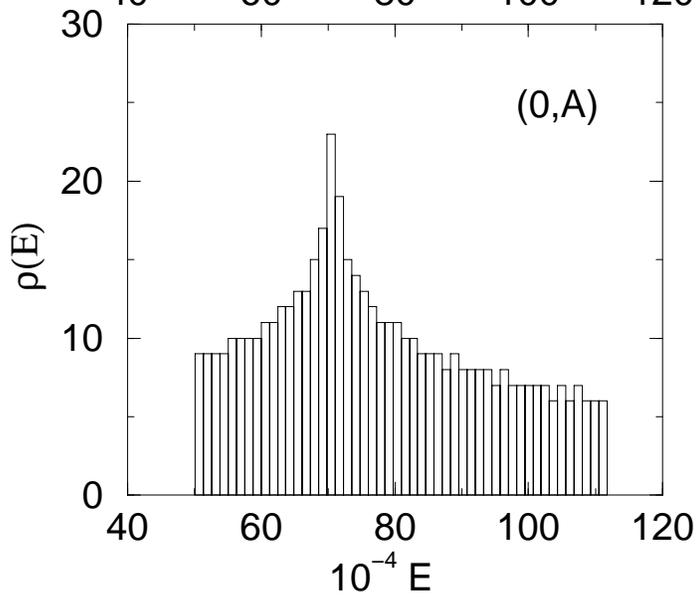
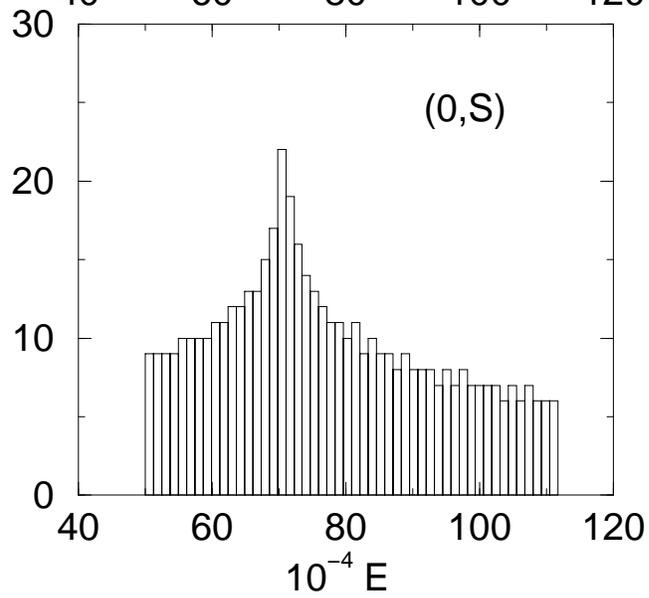
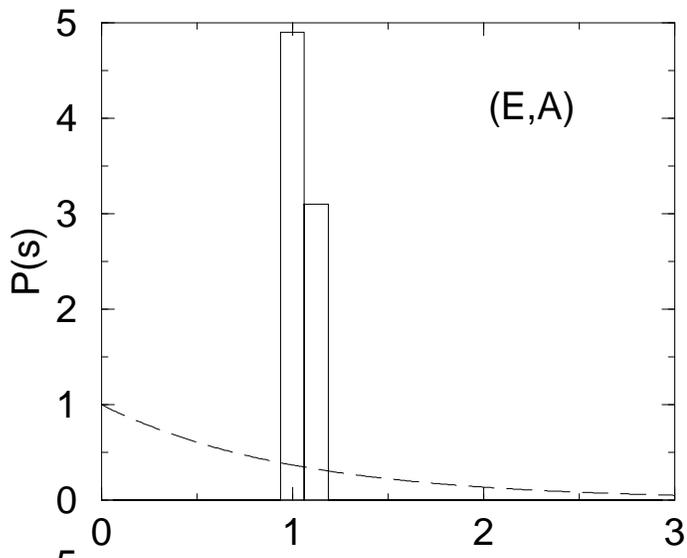
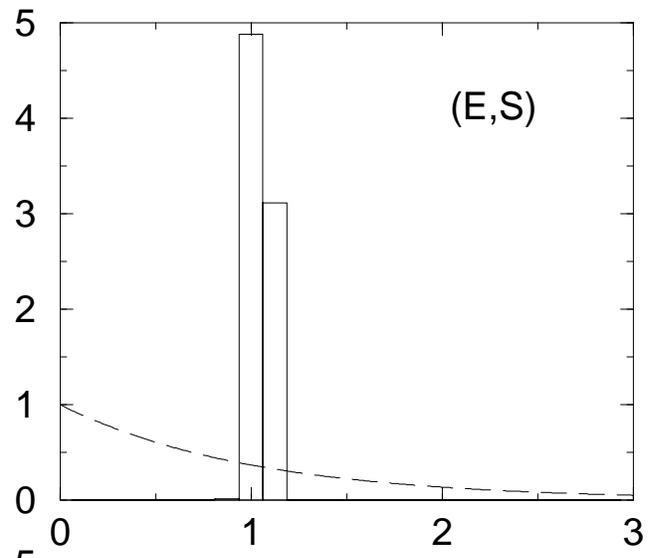
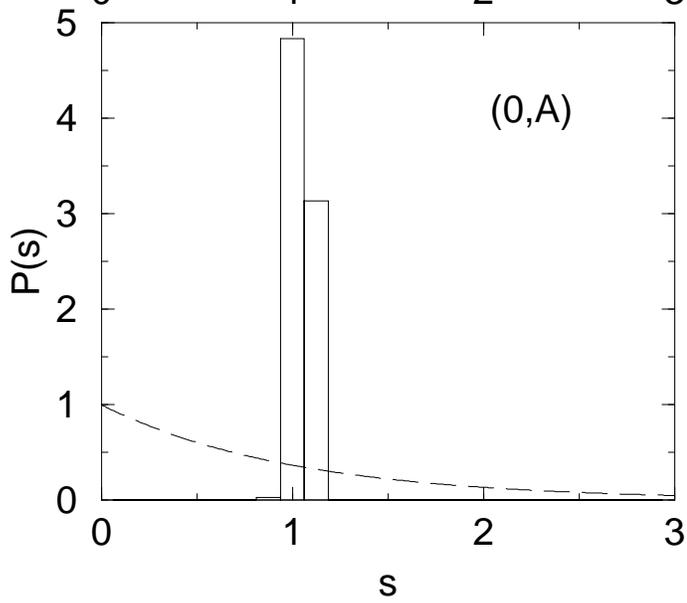
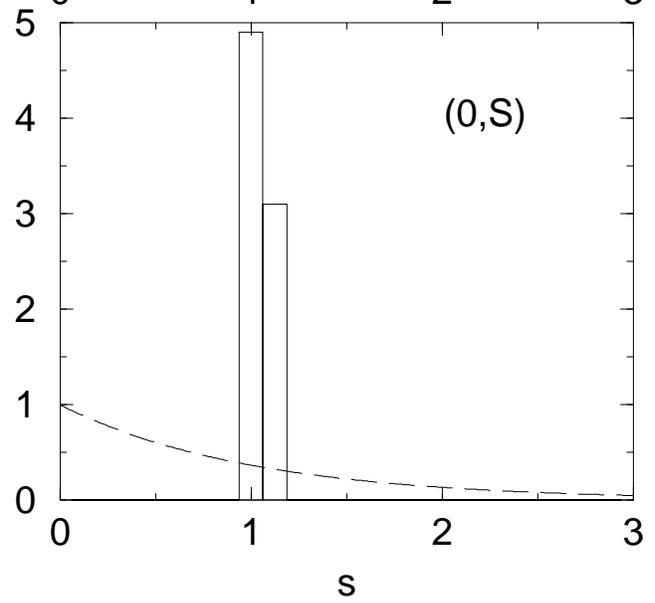
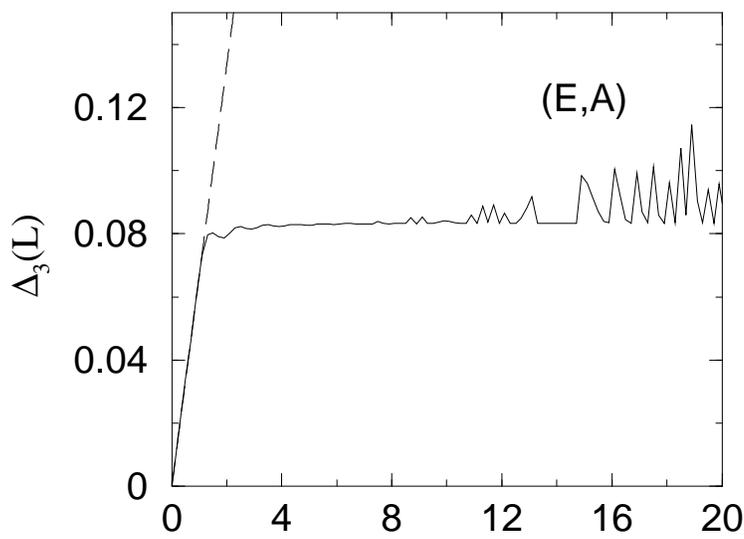
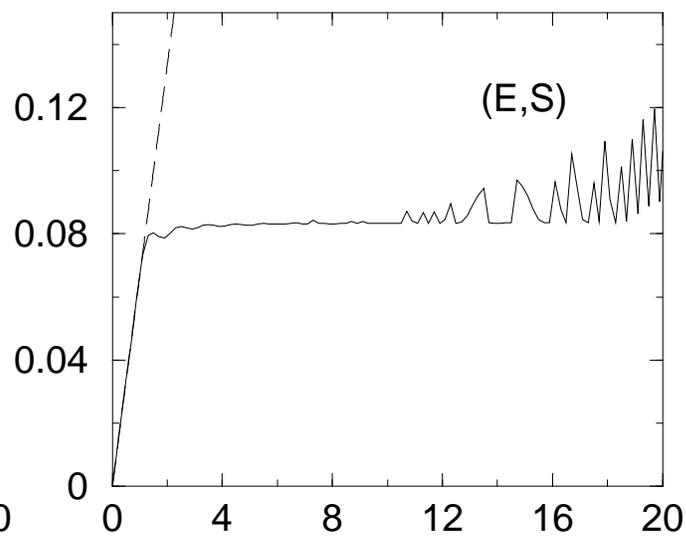
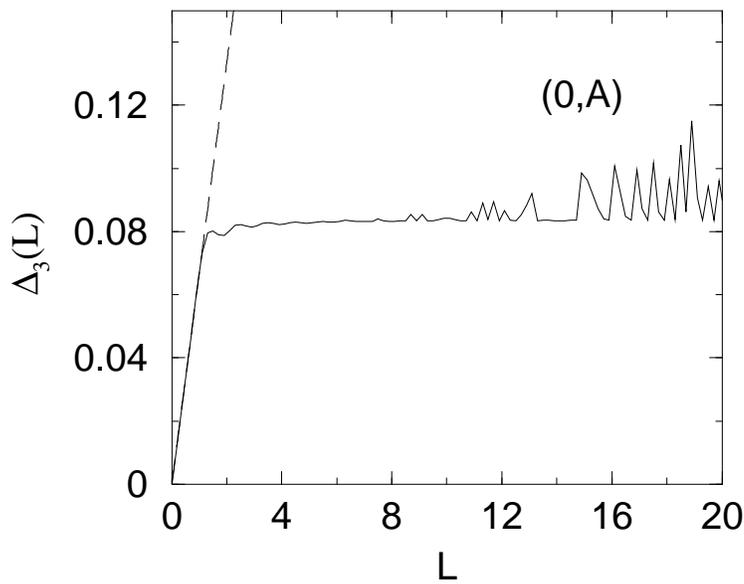
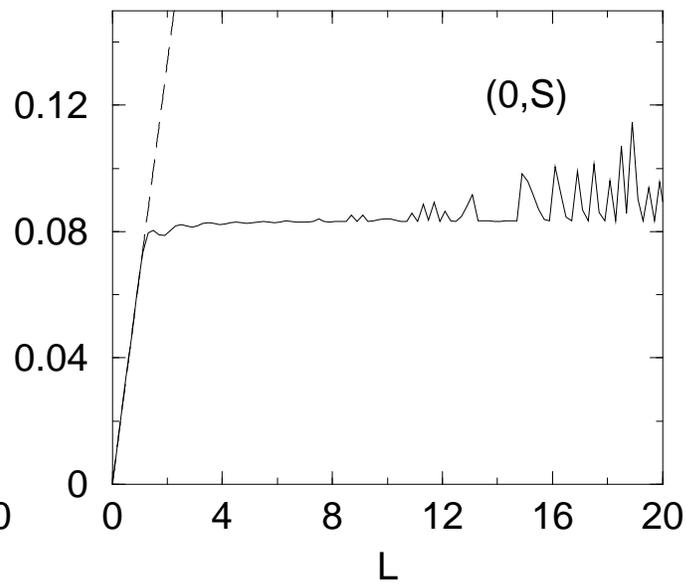
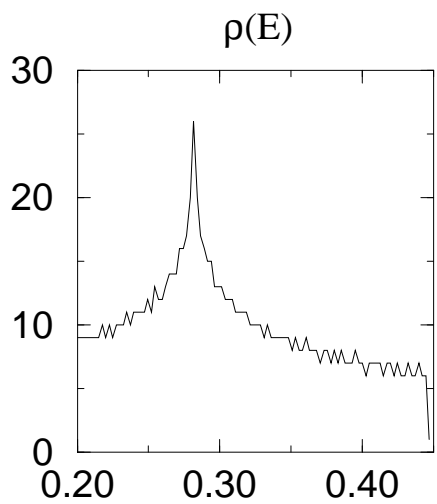
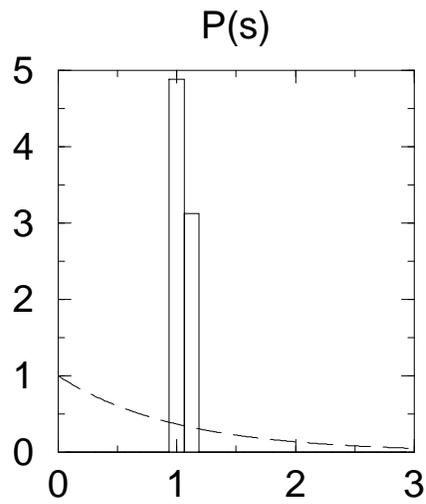
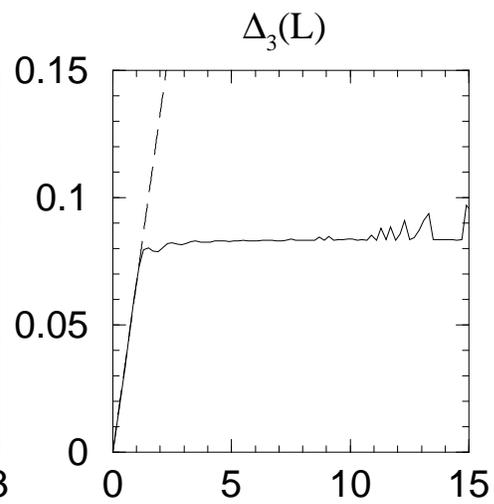
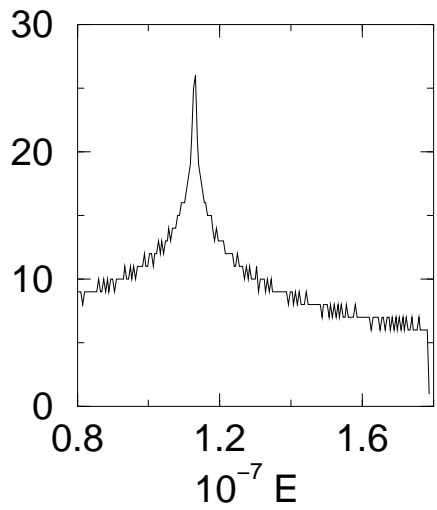
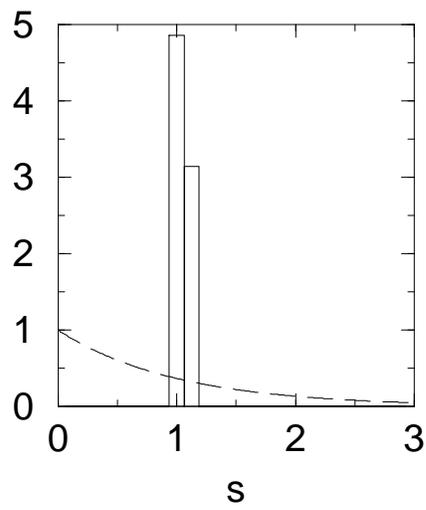
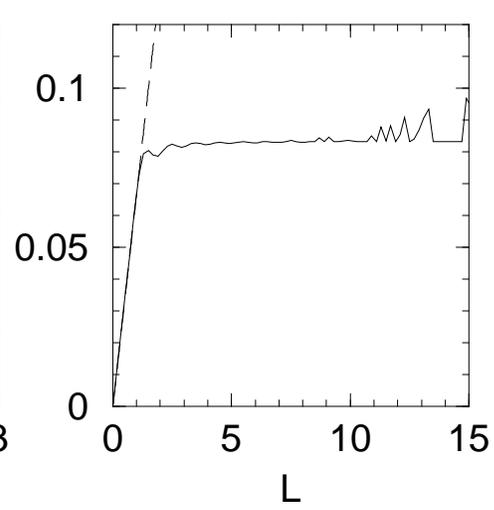
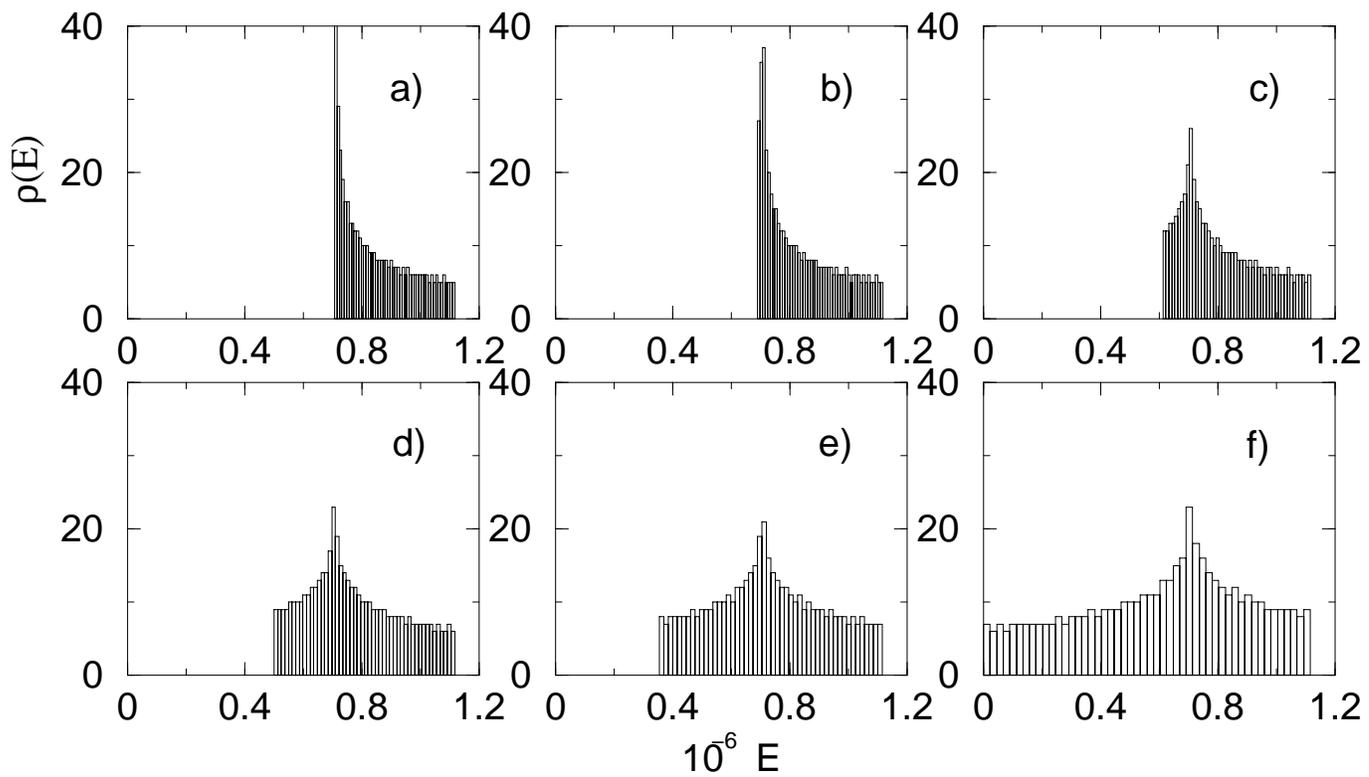
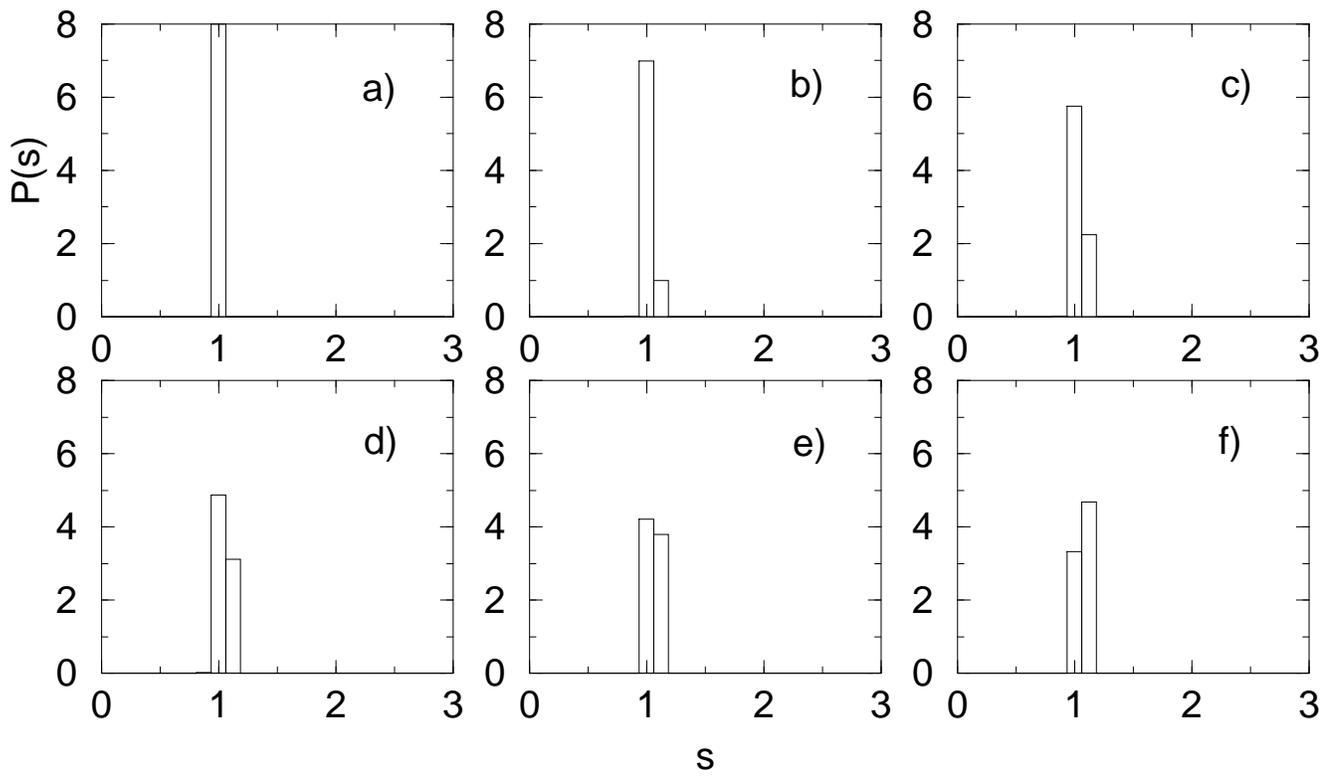
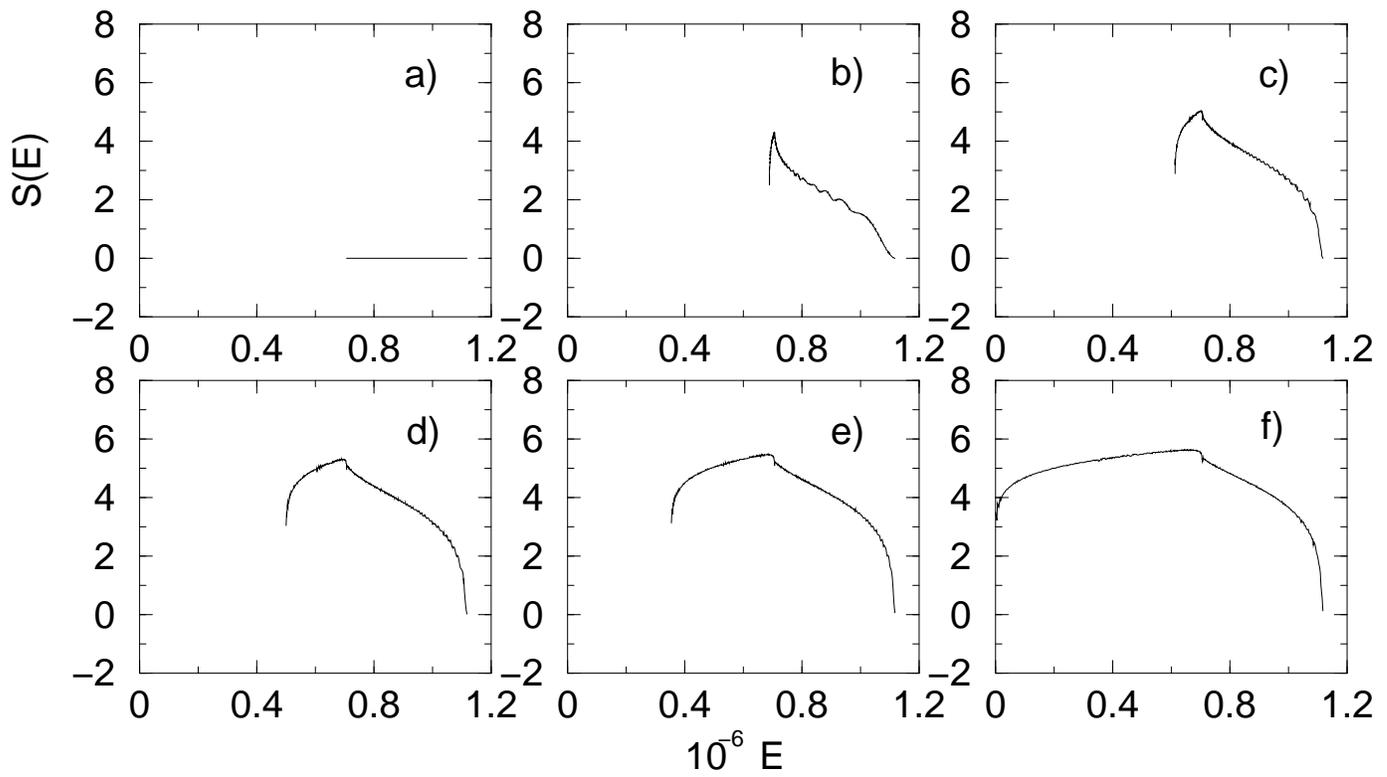
\end{document}